\documentclass[sigconf]{acmart} % ACM template (two-column)
\AtBeginDocument{%
  }

\setcopyright{acmlicensed}
\copyrightyear{2026}
\acmYear{2026}
\acmDOI{}
\acmISBN{}
\acmPrice{}
\acmConference[Healthcare Beyond Reaction Workshop @ IH '26]{ACM Interactive Health 2026}{July 05--08, 2026}{Porto, Portugal}
\acmBooktitle{\textit{Healthcare Beyond Reaction: Harnessing AI and Sensing for Proactive Care} Workshop at ACM Interactive Health 2026 (IH '26), July 05--08, 2026, Porto, Portugal}

\newcommand{\papertitle}{Exploring Reinforcement Learning for Fluid Transitions Between Clinical Mental Healthcare and Everyday Wellness Support}
\newcommand{\papershorttitle}{Exploring RL for Fluid Clinical-Wellness Care Transitions}
\newcommand{\paperkeywords}{reinforcement learning, digital mental health, adaptive interventions, journaling, contextual bandits, care continuity} % Separate the keywords with commas. No period at the end.

\usepackage{enumitem}

\usepackage{xcolor}
\usepackage{multirow}
\usepackage{colortbl} % for coloring select table rows
\usepackage{array}
\usepackage{siunitx}
\usepackage{graphicx}
\usepackage{dblfloatfix}
\usepackage{xspace}
\usepackage{hyperref} % enable href in footnotes, load hyperref last
\usepackage{tabularx} % If tabularx, load it after hyperref

\newcommand{\para}[1]{\vspace{0.2em}\noindent\textbf{\textit{#1}}\hspace*{.3em}}

\usepackage[backgroundcolor=white, textcolor=magenta, linecolor=magenta, bordercolor=white, textsize=tiny]{todonotes}
\newcommand{\placeholder}[1]{\textcolor{gray}{(#1)}} 

\usepackage{lipsum}
\newcommand{\placeholderPARA}[1][]{
\ifthenelse{\equal{#1}{}}{\placeholder{\lipsum[2]}}{\placeholder{\lipsum[1][1-#1]}}
}

\newcommand{\deletes}[1]{}

\begin{document}

\title[\papershorttitle]{\papertitle}

\author{Tony Wang}
\email{yw2567@cornell.edu}
\orcid{0000-0002-6074-3127}
\affiliation{%
  \institution{Cornell University}
  \country{United States}
}

\author{Qian Yang}
\email{qianyang@cornell.edu}
\orcid{0000-0002-3548-2535}
\affiliation{%
  \institution{Cornell University}
  \country{United States}
}

\renewcommand{\shortauthors}{Wang and Yang}

\begin{abstract}
    % IH WS V2 (word limit: 170 words)
Mental health struggles wax and wane, yet clinical and wellness interventions typically operate separately, causing frequent breakdowns at care transitions. We explore reinforcement learning (RL) as a means to build digital health systems that deliver clinical and wellness interventions proactively, as part of a coherent care journey. We ask: what complexities does designing such a system involve? 
We built a contextual bandit that dynamically selects journaling prompts from clinical and wellness repertoires to optimize for an overarching health goal (sustained journaling) and deployed it in a four-week exploratory study (N=38).
We found that, first, many benefits of RL-optimized intervention sequences appeared only after interventions ended, raising the question: \textit{Should systems that offer coherent clinical-wellness care journeys include stepping-back periods? If so, when and how?}
Second, participants most engaged with RL-generated interventions deepened their engagement over time, while those most engaged with a constant intervention tended to burn out and drop out later. It raises the question: \textit{when should a system blending clinical and wellness interventions reduce intensity to prevent burnout in versus sustain it to maximize treatment gains?}
\end{abstract}

%%
%% The code below is generated by the tool at http://dl.acm.org/ccs.cfm. Update it as you need.
%%
\begin{CCSXML}
<ccs2012>
   <concept>
       <concept_id>10003120.10003123.10011759</concept_id>
       <concept_desc>Human-centered computing~Empirical studies in interaction design</concept_desc>
       <concept_significance>500</concept_significance>
       </concept>
   <concept>
       <concept_id>10010147.10010257.10010293</concept_id>
       <concept_desc>Computing methodologies~Machine learning approaches</concept_desc>
       <concept_significance>300</concept_significance>
       </concept>
 </ccs2012>
\end{CCSXML}

\ccsdesc[500]{Human-centered computing~Empirical studies in interaction design}
\ccsdesc[300]{Computing methodologies~Machine learning approaches}

\keywords{\paperkeywords}

% \input{fig_table/fig_teaser}

\maketitle

\section{Introduction}

% P1 [Overarching hero]: interventions of various intensity adapt fluidly to mental health/well-being fluctuations
Mental health struggles wax and wane, requiring interventions that vary in intensity and adapt fluidly to them.
Anxiety disorders cycle through acute flares and partial remissions~\cite{angst2009generalized}.
Depression recurs in over half of those who recover from a first episode~\cite{malhi2018depression, wittchen2000waxing}. 
Digital health systems that sense these shifting needs and blend clinical interventions with everyday wellness support into a coherent care journey could intervene before a first episode or full relapse~\cite{mohr2017personal,rubanovich2017health} and, during periods of relative stability, sustain benefits from clinical treatment~\cite{nahum2016just}.

% P2 [Overarching villain] Silos. Interventions of various intensity delivered seperately.
This vision of a coherent care journey is not yet a reality: systems that deliver more intense mental health interventions from clinical repertoires (e.g., trauma-focused grounding techniques) typically operate separately from systems that provide everyday well-being support (e.g., sleep, exercise, journaling)~\cite{YangGangCHI24_MAMH,lamonica2022informing,YangGangCHI26_NutritionalSupplements,najavits2002seeking}.
Even when the same person uses both for the same condition, the two share no data, align to different goals and timelines, and offer no cross-referral when care needs intensify or ease~\cite{bauer2020smartphones,iorfino2021using, YangGangCHI24_MAMH}.
This fragmentation hinders proactive and timely care, creating many harms: frequent breakdowns at care transition points~\cite{Chen2024Digital,Backman2024Platform}, delayed care escalation~\cite{YangGangCHI24_MAMH}, prolonged unnecessary treatment~\cite{DanAdlerCHI22_PsychiatricDrugDiscontinuation,poolen2025systematic,douven2021payment}, and measurably poorer health outcomes~\cite{Prior2023Healthcare,Backman2024Platform,lamonica2022informing}.

% P3: Offer our solution framework + situate ''tech that delivers interventions of various intensity as one coherent journey'' within it
We envision a future where digital health systems proactively identify users' changing needs and flexibly deliver interventions from both clinical and wellness repertoires as a coherent care journey.
Achieving this vision requires coordinated efforts across many fronts, several of which various research communities are already pursuing.
For instance, it requires changes to business incentives and regulatory policies so that wellness technologies augment rather than delay or replace necessary clinical care~\cite{YangGangCHI26_NutritionalSupplements,simon2022skating}.
It calls for novel system designs that blend clinical and wellness interventions effectively while preserving clinician authority over treatment decisions and patient choice in everyday well-being support.
It also requires redesigned clinical workflows that incorporate data from both sides to coordinate care~\cite{cohen2016integrating,nghiem2023understanding}. 

As HCI researchers, we have been contributing on the policy~\cite{YangGangCHI26_NutritionalSupplements,YangGangCHI24_MAMH} and technology design fronts~\cite{YangGangDIS25_Identity_writing}, with a particular focus on NLP-enabled mental health systems.
We bring this broader work to the workshop, while this paper presents one of our ongoing projects on the technology design front.

\section{Motivation}
% P4: [SH] A new approach: RL-powered digital health platform that detects user situations and delivers interventions of various intensity as one coherent journey
This paper explores reinforcement learning (RL) as a means to blend clinical mental health care and wellness support into a coherent care journey.
RL optimizes a \textit{sequence} of recommendations toward a cumulative long-term goal, re-evaluating at every step as the user changes themselves and as each previous recommendation changes the user~\cite{yu2021reinforcement,tewari-mobilehealth2017-adstointerventions,murphy2003optimal}.
These capabilities make RL a particularly promising candidate for our goal:

\begin{itemize}[leftmargin=*]
    \item \textit{RL could dynamically adapt intervention choices to users' mental health fluctuations, while steering toward long-term health goals~\cite{aguilera2024effectiveness,daskalova-chi2021-selfe,yom2017encouraging}.}
    % Prior work has applied RL to delivering either clinical interventions~\cite{aguilera2024effectiveness} or wellness support~\cite{daskalova-huang-chi2020-sleepbandits,yom2017encouraging}, never both.
    \item \textit{With more granular intervention options, RL could enable more granular transitions between clinical and wellness support.~}While stepped care moves patients between fixed tiers of treatment intensity~\cite{bower2005stepped}, RL could make fine-grained intervention decisions, for example, by mixing clinical and wellness micro-interventions in varying ways over time. An RL journaling system, for instance, could recommend grounding exercises from trauma therapy one week, then a how-was-my-day prompt the next day.
    \item \textit{RL could allow clinicians, users, and the model to jointly coordinate interventions from both repertoires toward shared health goals.~}Existing care transition mechanisms, such as stepped care~\cite{bower2005stepped} and patient-clinician data sharing~\cite{cohen2016integrating, bauer2020smartphones}, place transition decisions in the hands of one party. RL distributes these decisions: clinicians and users together define care goals and curate the micro-intervention pool, while the model autonomously selects and sequences interventions toward that shared goal.
\end{itemize}

% P5: [SV] The unknown unknowns about designing these systems, beyond just the effectiveness of this approach.
Despite these promises, designing a single RL system that blends interventions from clinical and wellness repertoires remains largely unexplored.
Evaluating whether such systems improve health outcomes is both important and challenging~\cite{cuijpers2019role,kazdin2007mediators}, but before we can evaluate, we need to first ask: \textit{what design complexities and pitfalls emerge when a system operates across the clinical-wellness divide? }Without surfacing these ``\textit{unknown unknowns},'' we risk building systems with avoidable design flaws and overlooking unintended consequences of operating across this divide.

% P6: Research Goal/RQ
This research takes an initial step toward surfacing these unknown unknowns.

\section{Method}
% P1: Overarching approach: (1) building a simple RL model, (2) sustained jouranling as goal, (3) deploy (not "evaluate") with generally-healthy users only (not clinicians)
As an initial exploration, we built a simple RL system that optimizes toward \textit{sustained journaling} as its overarching health goal, and deployed it with a small group of generally healthy crowd-workers.
Several considerations led us to choose this approach.

% P1: (2) why sustained jouranling as goal
We chose \textit{sustained journaling} as the health goal our RL system optimizes toward, because blending clinical and wellness interventions for this goal is both helpful and achievable.
Helpful, because clinical journaling prompts can address mental health struggles, while wellness prompts improve everyday well-being; together they can benefit the same person as their situation changes~\cite{sohal2022efficacy,smyth2018online,emmons2003counting,king2001health}.
Achievable, because interventions supporting journaling exist in both clinical~\cite{miller2014interactive} and wellness repertoires~\cite{emmons2003counting}.

% P1: (2) Why building a simple RL model
We chose to build a simple RL model for this initial exploration, because our goal is to discover the complexities around designing systems that span clinical and wellness interventions, not to maximize model performance or health/well-being efficacy (yet).

% P4: (3) Why deploying (not "evaluate") with generally-healthy users only (not clinical population, not in clinical context)
We chose to deploy the RL system with a small group of generally healthy users, because (1) the interventions (journaling prompts) are generally low-risk enough\footnotemark[1] to deploy without clinical supervision, yet (2) the system wrapping them is untested enough that we wanted to test it with a lower-risk population first.

\footnotetext[1]{We intentionally selected only low-risk clinical interventions, ensuring they are appropriate for generally healthy individuals. For example, from trauma therapy, we selected only present-focused prompts (e.g., grounding exercises~\cite{najavits2002seeking}), not trauma-processing ones.}

\subsection{System Design}

% P1: Diverse yet low-risk prompts
We began by curating a diverse pool of proven journaling prompts from both clinical and wellness literature.
This resulted in $247$ journaling prompts, including:
\begin{itemize}[leftmargin=*]
    \item \textit{Prompts from the clinical repertoire:~}For example, $51$ grounding exercises from trauma and anxiety therapy\footnotemark[1]~\cite{najavits2002seeking}, $8$ cognitive reframing prompts from cognitive behavioral therapy~\cite{beck2024cognitive}.
    \item \textit{Prompts from the wellness repertoire:~}For example, $21$ gratitude journaling prompts~\cite{emmons2003counting}, $6$ best-possible-self prompts from Positive Psychology Intervention (PPI)~\cite{king-issuesineducation2006-ejournaling}. % $75$ creative writing prompts~\todocite, 
    \item \textit{Prompts that straddle both:~}For example, $4$ expressive writing prompts, which originate in clinical trauma research~\cite{pennebaker-journalabpsych1986-expressivewritingtrauma} but are now widely practiced for general well-being.
\end{itemize}

% P2: Why this model + Why this reward signal
We trained a simple contextual bandit to select journaling prompts from this pool, using journal entry length (word count) as its reward signal.
Word count is by no means a perfect proxy for \textit{sustained journaling}, but it offers several advantages: it is objective, it does not favor one journaling prompt repertoire over the other, and it is directly observable, allowing the bandit to learn automatically after every journaling session.
The bandit uses an epsilon-greedy strategy, exploiting the highest-reward prompt 80\% of the time and exploring randomly 20\%.

% P4: Personalization features
To predict which prompts will maximize a user's journal entry length the next day, the bandit draws on six context features: four from the prior session (insertions per minute, deletions per minute, self-reported writing difficulty, self-reported prompt difficulty) and two from registration (prior journaling experience, prior mental health service use).
Each participant's first 3--5 prompts (the cold-start period) are selected randomly.

We embedded the bandit and prompt pool into \textsc{CoAuthor}~\cite{lee2022coauthor}, a research platform for studying human-AI collaborative writing.
The resulting system, \textsc{LongVue}, presents users with a daily journaling prompt selected by the bandit (or other models of researchers' choosing), captures keystroke-level writing behavior, and records post-writing survey responses.
 % Key-stroke level logging also helped us screen participants who copy-pasted texts from generative AI tools.

\subsection{User Study Design}

% P1: Conditions
We deployed the system in a user study with three conditions: journaling with RL-selected prompt sequences, randomly selected prompts, and a consistent prompt.
The random condition tests whether RL optimization adds value beyond prompt variety alone, while the consistent condition (\textit{``What's on your mind today?''}) isolates the effect of variety itself.

% P2: Temporal structure
The study lasted four weeks. In the first two weeks (\textit{``intervention''}), participants journaled daily with prompts from their assigned condition. During the following two weeks (\textit{``post-intervention''}), participants journaled freely with no prompts, with lowered participation requirements.
The post-intervention stage allows us to observe the gap between what the RL optimizes for (next-day journal entry length) and the health goal it approximates (sustained journaling).

% P3: Data collected
We collected three types of data from this process: behavioral data from each session (word count, time spent, keystroke dynamics such as insertion and deletion patterns); post-session surveys (self-reported writing difficulty and prompt difficulty on 5-point scales); and the journal entries themselves. We also administered a brief exit survey about participants' overall experience.

\begin{table}[b]
  \begin{tabular*}{\columnwidth}{l @{\extracolsep{\fill}} ccc}
    \toprule
    & Consistent & Random & RL \\
    \midrule
    Participants Past Screening
      & 12 & 13 & 13 \\
    Dropped during intervention 
      & 5* & 5 & 6 \\
    Dropped post-intervention 
      & 2 & 1 & 1 \\
    Completed both stages 
      & 5 & 7 & 6 \\
    \bottomrule
  \end{tabular*}
  \vspace{0.1cm}
  
  {*: Includes 3 intervention-related dropouts.}
  \vspace{0.2cm}
  \caption{Participant flow across study stages by condition. Collectively, they produced 648 legitimate journal entries (e.g., not copied from elsewhere according to system log.)}
  % 648 is the number post CSCW quality screening (excl. those copy pasted from elsewhere, empty entries, <100 words.)
  % Some participants dropped out not due to attrition but as a meaningful signal of the intervention's effects. Section~\ref{sec:data_analysis} details how we carefully identified these intervention-linked dropouts.
  \label{tab:dropout}
  \vspace{-0.5cm}
\end{table}
\begin{figure*}[htb]
\centering
\includegraphics[clip, width=\linewidth]{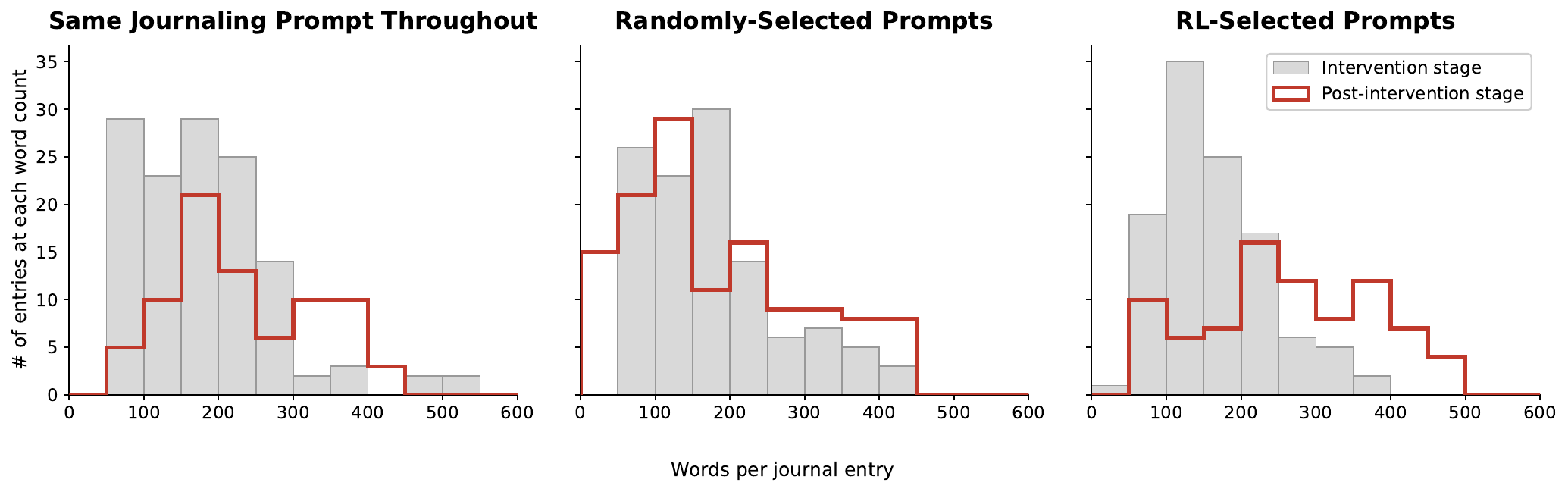}
\vspace{-0.7cm}
\caption{Participants who received RL-sequenced prompts from clinical and well-being repertoires wrote longer entries only after those prompts ended (right panel, right half of the red box in the right panel taller than that of other panels), an observation much less pronounced in the other two conditions.} % Journaling engagement (proxied by word count) during intervention (gray bars) and post-intervention (red outline) for participants who completed both stages. % Randomly-selected prompts from the same pool showed a smaller effect (middle panel). The same prompt throughout showed no change (left panel). 
\Description{Participants who received RL-sequenced prompts from clinical and well-being repertoires wrote longer entries only after those prompts ended (right panel, right half of the red box in the right panel taller than that of other panels), an observation much less pronounced in the other two conditions.}
\label{fig:post_intervention}
\end{figure*}

\para{Participants.~}
% P1: Who, recruitment, assignment
We recruited generally healthy crowd-workers from Prolific. Of $48$ who responded, $38$ passed initial screening and were randomly assigned in equal numbers to the three conditions.

% P2: Compensation structure + why designed this way
Because compensation strongly affects engagement in a paid study, we designed the compensation structure carefully and tailored our analysis accordingly.
\begin{itemize}[leftmargin=*]
    \item \textit{During the intervention stage}, we incentivized participants to journal consistently, maximizing the effect of the intervention. They received \$30 for journaling on at least 11 of 14 days, regardless of entry length.
    \item \textit{Post-intervention}, we provided no incentive to write beyond the bare minimum. Participants needed only 4 journal entries over two weeks to earn an additional \$30, regardless of entry length.
\end{itemize}

\para{Data Analysis.~}\label{sec:data_analysis}
% P5: Valid signals given compensation (the list)
Our analysis focused on signals that can answer our research question and are unaffected by compensation:
\begin{itemize}[leftmargin=*]
    \item \textit{Intervention stage:} (i) cross-condition comparisons of engagement metrics (e.g., word count, typing speed), since all conditions received identical incentives; (ii) within-condition engagement variation tied to specific prompt events (e.g., word count changes when a prompt repeats), since compensation incentivizes frequency, not entry length; (iii) journal content and participants' perceptions of individual prompts (e.g., daily survey responses), when they explain mechanisms behind an already-observed engagement pattern.
    \item \textit{Post-intervention stage:} (i) cross-condition comparisons of engagement levels, since all conditions shared the same minimal requirement; (ii) cross-condition comparisons of how engagement changes between stages, since all conditions shared the same incentive structure; (iii) participants' perceptions of the journaling experience (e.g., exit interviews).
\end{itemize}

% P6: Dropout handling
Notably, we treated dropout as meaningful \textit{only when a participant's writing or survey response before they left connects their disengagement to the intervention mechanism} (e.g., escalating emotional depth followed by abrupt disengagement, or declining word counts after receiving a repeated prompt). This allows us to differentiate inevitable attrition common to any crowd-worker study, from dropout that reflects the intervention's impact on a participant's sustained journaling practice (Table~\ref{tab:dropout}).

% P7: Analytical approach (descriptive, qualitative, small N)
Given the modest sample sizes, we focus on descriptive patterns and qualitative analysis rather than inferential statistics.

% P4 defends the analytical approach: "Why not run statistical tests?" Because with 12-13 per condition, the goal is surfacing patterns, not confirming effects.
% We analyzed engagement patterns across conditions and stages using both behavioral metrics and qualitative examination of writing content. Primary metrics included word count, time spent writing, and keystroke dynamics (insertions and deletions per minute). We compared participants who continued to the unprompted stage against those who dropped out, and qualitatively examined writing logs for patterns in content that might explain the behavioral differences.

\section{Finding 1: RL's Benefits Post-Intervention}

% P0: Finding overview
Our findings are twofold.
First, the benefits of RL optimization appeared only after the intervention ended.

% P1: Observation #1: During intervention, RL participants didn't write longer than other conditions + complained about the prompt repetition.
During the intervention stage, RL-condition participants were no more engaged with the journaling task than other participants.
They spent similar time and wrote a similar amount of words as participants writing with the same prompt every day, and fewer words than the random-condition participant (median entry length: 208 words for RL, 203 for static, 231 for random).
Several RL-condition participants even complained about prompts recurring, a consequence of the RL model's exploitation behavior, and their journal length dropped on those days.

% P2: Observation #2: After prompts ended, RL participants wrote notably longer entries than during intervention + recalling & appreciating prompts.
Post-intervention, however, RL-condition participants became notably more engaged. They wrote substantially longer entries, increasing from the 200–300 word range to the 300–450 word range (Figure~\ref{fig:post_intervention}).
Their writing also grew in reflectivity. On day 21, for example, one RL participant devoted an entire entry to reflecting on their intervention-stage experience, identifying which prompts were hardest, which were most memorable, and appreciating the diversity of the prompt sequence, noting the mix of ``\textit{typical journaling prompts and some more creative writing type ones}''.

% P3: These observations are genuine signals of benefit to health outcome.
These post-intervention changes reflect genuine progress toward sustained journaling, not compensation effects, because participants received no rewards for writing longer or more frequently during this phase.

% P4: This benefit is attributed to RL, not randomization.
These post-intervention changes only appeared in the RL condition.
The random condition, which drew from the same prompt pool without optimization, showed no comparable post-intervention increase (median word count: 208 during intervention to 192 post-intervention; mean: 231 to 232; Figure~\ref{fig:post_intervention}, middle panel). The static condition showed no post-intervention change either (left panel).

\para{Implication for Designing Digital Health Systems Across Clinical and Wellbeing Lines.~}
This finding echoes a phenomenon known as the ``\textit{sleeper effect}'' in psychotherapy, where therapeutic gains emerge or strengthen after treatment ends, sometimes exceeding gains made during treatment itself~\cite{shedler2010efficacy, fluckiger2017sleeper}.
Researchers attribute this to several mechanisms, for example, that skills acquired during therapy consolidate only through independent application without therapist scaffolding~\cite{hollon2006enduring}.
Our intervention is far lighter-touch than clinical psychotherapy, yet a similar consolidation pattern appeared.

% This paragraph: Future designs should design stepping-back period. 
Through this lens, this finding suggests that digital health systems spanning clinical and wellness care may benefit from deliberate stepping-back periods. Because fragmented, episodic care~\cite{YangGangCHI24_MAMH,YangGangCHI26_NutritionalSupplements} is a core problem these systems aim to solve, a natural instinct might be to provide continuous, ``\textit{everyday}'' interventions. But in our study, RL's benefits appeared only after the system stopped, suggesting that in a coherent care journey, when to step back is itself an important design decision, not necessarily a problem, nor merely an absence of intervention.

We argue that how to design these stepping-back periods when interventions span clinical and wellness repertoires is an important, open research question. Stepping back carries different consequences for clinical and wellness interventions. Pausing a wellness intervention, such as a gratitude prompt, primarily costs a missed opportunity for well-being maintenance. By contrast, pausing a clinical intervention, such as a grounding exercise for anxiety management, risks leaving a user without needed therapeutic support. A system that delivers only one kind of intervention does not need to navigate this tension: \textit{when a system delivers both, how should it weigh the consolidation benefits of stepping back against the risks of withdrawing needed clinical support?}

% \newpage
\section{Finding 2: The Paradox of High Engagement}

% P1: Among highly engaged participants, outcomes diverged across conditions.
This finding concerns participants who, compared to others, were highly engaged during the intervention stage.
Such participants in the RL condition (and to a lesser extent,  in random condition) tended to stay and become even more engaged post-intervention, while those in the consistent-prompt condition tended to drop out.

% P2: [Pathway #1] In RL and random, the most engaged participants tended to persist.
In both the RL and random conditions, the most engaged participants during the intervention stage tended to be the ones who persisted .
In the RL condition, participants who eventually journaled for the whole four weeks had written more than twice as many words in the intervention stage as those who later dropped out (242 vs.\ 105 words per entry). In the random condition, completers had written roughly a quarter more than dropouts (243 vs.\ 195 words).

% P3: [Pathway #2] In the static condition, this pattern reversed. The most engaged tended to drop out. 
In the consistent-prompt condition, this pattern reversed: the most engaged participants during the intervention stage tended to be the ones who dropped out.
Before dropping out, they had written more than those who eventually completed the study (237 vs. 196 words per entry), typed at a higher rate (206 vs. 173 insertions per minute), and revised far less (8.9 vs. 16.5 deletions per minute).

% P4: Qualitative examination of these dropouts' writing suggests a possible mechanism.
Our qualitative analyses of several of these participants' writing might explain why: it grew increasingly personal and emotional before they abruptly stopped.
Of the five consistent-condition participants who dropped during the intervention, three had been writing about deeply personal topics (e.g., a parent's dementia, a woman they recently fell in love with), with high volume, fast typing, and minimal revision (Table~\ref{fig:post_intervention}).
Those who completed the same condition did not write similarly personal content. As one put it: ``\textit{My life is too steady, no drama\ldots{} [Journaling] didn't make a difference in how I felt about it.}''

\para{Implication for Designing Digital Health Systems Across Clinical and Wellbeing Lines.~}
% P1: Tie finding to RW
The forking paths of highly engaged participants that we observed appear to echo the two dynamics documented in clinical and wellness literature, respectively.
Highly engaged participants persisting in the varied-prompt conditions are consistent with behavior change research, where variety sustains engagement and monotony breeds disengagement~\cite{kovacs2021not}. Highly engaged participants writing increasingly personal and emotional content before dropping out is consistent with clinical research, where unstructured emotional processing without redirection can become counterproductive, particularly for those prone to rumination~\cite{sbarra2013expressive}.

% P2: This is unique (claim this softly) to cross-repertorie systems.
Because a system blending clinical and wellness interventions draws from both repertoires, both dynamics can co-occur, creating a challenge that neither line of research has faced alone. Behavior change research addresses habituation by varying intervention content to sustain engagement. Clinical practice addresses emotional depth through clinician oversight, redirecting when emotional processing becomes counterproductive. Neither has had to manage both dynamics within the same system.

% P3: Future work
\textit{For its most engaged users, when should a system that offers both clinical and wellness interventions reduce their engagement to prevent burnout, versus sustain it to maximize health gains?}
Prior research in clinical and health habit formation fields has addressed each dynamic separately: variety prevents habituation, and clinician oversight manages emotional depth. But the question of how to navigate both within one system has remained underexplored.
Yet even in our simple deployment, this tension had real consequences: the most engaged users tended to either deepen their practice or disengage entirely. 
In this light, we see a ready opportunity for future research to explore how systems that blend clinical and wellness interventions should navigate this tension for their most engaged users, a design challenge that arises when both repertoires meet in one system.

% \section{Closing Note}
% This work takes a moderate first step towards digital health systems that deliver interventions from both clinical and wellness repertoires as one coherent care journey. Our simple, initial exploration surfaced two design tensions underexplored in neither tradition faces alone, and merit further study.
% when to design stepping-back periods that allow consolidation without premature withdrawal, and how to read the forking paths of highly engaged users who may be deepening benefit or approaching disengagement.

% Limitation
% The sample is small (5--7 per group in the free-writing stage), the bandit condition ran after the random condition (a temporal confound), and all participants were paid crowdworkers; we cannot determine whether the qualitative contrast caused the quantitative divergence.
%% The acknowledgments section is defined using the "acks" environment (and NOT an unnumbered section). This ensures the proper

\begin{acks}
Research reported in this publication was supported by the National Library of Medicine of the National Institutes of Health under award number 1R01LM014306-01. The ideas represented here were partially developed during the 2025 Everyday AI and Mental Healthcare Thought Summit, sponsored by the Cornell Center for Data Science for Enterprise and Society. 
\end{acks}

% This material is also based upon work supported by the National Science Foundation under Grant No. 2313078 (HCC Social Media TestDrive). Any opinions, findings, and conclusions or recommendations expressed in this material are those of the author(s) and do not necessarily reflect the views of the National Science Foundation.

\bibliographystyle{ACM-Reference-Format}
\bibliography{ref/misc,ref/writing, ref/identity, ref/YangGangPrior2025,ref/mentalhealth,ref/journaling}

@inproceedings{YangGangCHI26_NutritionalSupplements,
    author = {Cooper, Ned and Guridi, Jose A. and Hwang, Angel Hsing-Chi and Kolko, Beth and McGinty, Emma Elizabeth and Yang, Qian},
    title = {Framing Responsible Design of AI for Mental Well-Being: AI as Primary Care, Nutritional Supplement, or Yoga Instructor?},
    year = {2026},
    isbn = {9798400722783},
    publisher = {Association for Computing Machinery},
    address = {New York, NY, USA},
    url = {https://doi.org/10.1145/3772318.3791556},
    doi = {10.1145/3772318.3791556},
    booktitle = {Proceedings of the 2026 CHI Conference on Human Factors in Computing Systems},
    articleno = {1441},
    numpages = {16},
    location = {Barcelona, Spain},
    series = {CHI '26}
}

@article{YangGangDIS25_Identity_writing,
    author = {Wise, Talia and Yang, Yuewen and Shim, Ryun and Chang, Kevin Chuan-Kai and Oden Choi, Judeth and Yang, Qian},
    title = {Investigating How Emerging Adults Explore Identity through Writing: Opportunities for AI Writing Assistants to Help},
    year = {2025},
    isbn = {9798400714856},
    publisher = {Association for Computing Machinery},
    address = {New York, NY, USA},
    url = {https://doi.org/10.1145/3715336.3735848},
    doi = {10.1145/3715336.3735848},
    booktitle = {Proceedings of the 2025 ACM Designing Interactive Systems Conference},
    pages = {2270–2282},
    numpages = {13},
    series = {DIS '25}
}

@inproceedings{YangGangCHI24_MAMH,
    author = {Hwang, Angel Hsing-Chi and Adler, Dan and Friedenberg, Meir and Yang, Qian},
    title = {Societal-Scale Human-AI Interaction Design? How Hospitals and Companies are Integrating Pervasive Sensing into Mental Healthcare},
    year = {2024},
    publisher = {Association for Computing Machinery},
    url = {https://doi.org/10.1145/3613904.3642793},
    doi = {10.1145/3613904.3642793},
    booktitle = {Proceedings of the 2024 CHI Conference on Human Factors in Computing Systems},
    series = {CHI '24}
}

@article{sohal2022efficacy,
  title={Efficacy of journaling in the management of mental illness: a systematic review and meta-analysis},
  author={Sohal, Monika and Singh, Pavneet and Dhillon, Bhupinder Singh and Gill, Harbir Singh},
  journal={Family medicine and community health},
  volume={10},
  number={1},
  pages={e001154},
  year={2022}
}

@article{smyth2018online,
  title={Online positive affect journaling in the improvement of mental distress and well-being in general medical patients with elevated anxiety symptoms: A preliminary randomized controlled trial},
  author={Smyth, Joshua M and Johnson, Jillian A and Auer, Brandon J and Lehman, Erik and Talamo, Giampaolo and Sciamanna, Christopher N},
  journal={JMIR mental health},
  volume={5},
  number={4},
  pages={e11290},
  year={2018},
  publisher={JMIR Publications Inc., Toronto, Canada}
}

@article{emmons2003counting,
  title={Counting blessings versus burdens: an experimental investigation of gratitude and subjective well-being in daily life.},
  author={Emmons, Robert A and McCullough, Michael E},
  journal={Journal of personality and social psychology},
  volume={84},
  number={2},
  pages={377},
  year={2003},
  publisher={American Psychological Association}
}

@article{king2001health,
  title={The health benefits of writing about life goals},
  author={King, Laura A},
  journal={Personality and social psychology bulletin},
  volume={27},
  number={7},
  pages={798--807},
  year={2001},
  publisher={Sage Publications Sage CA: Thousand Oaks, CA}
}

@article{miller2014interactive,
  title={Interactive journaling as a clinical tool},
  author={Miller, William R},
  journal={Journal of mental health counseling},
  volume={36},
  number={1},
  pages={31--42},
  year={2014},
  publisher={American Mental Health Counselors Association}
}

@article{wittchen2000waxing,
  title={The waxing and waning of mental disorders: evaluating the stability of syndromes of mental disorders in the population},
  author={Wittchen, Hans-Ulrich and Lieb, Roselind and Pfister, Hildegard and Schuster, Peter},
  journal={Comprehensive psychiatry},
  volume={41},
  number={2},
  pages={122--132},
  year={2000},
  publishers={Elsevier}
}

@article{angst2009generalized,
  title={The generalized anxiety spectrum: prevalence, onset, course and outcome},
  author={Angst, Jules and Gamma, Alex and Baldwin, David S and Ajdacic-Gross, Vladeta and R{\"o}ssler, Wulf},
  journal={European archives of psychiatry and clinical neuroscience},
  volume={259},
  number={1},
  pages={37--45},
  year={2009},
  publisher={Springer}
}

@article{malhi2018depression,
  title={Depression},
  author={Malhi, Gin S and Mann, J John},
  journal={The lancet},
  volume={392},
  number={10161},
  pages={2299--2312},
  year={2018},
  publisher={Elsevier Science Publishing Company, Inc.}
}

@article{Prior2023Healthcare,
  title={Healthcare fragmentation, multimorbidity, potentially inappropriate medication, and mortality: a Danish nationwide cohort study},
  author={Anders Prior and C. Vestergaard and P. Vedsted and Susan M. Smith and L. Virgilsen and L. Rasmussen and M. Fenger-Grøn},
  journal={BMC Medicine},
  year={2023},
  volume={21},
  doi={10.1186/s12916-023-03021-3}
}

@article{douven2021payment,
  title={Payment schemes and treatment responses after a demand shock in mental health care},
  author={Douven, Rudy and Remmerswaal, Minke and Vervliet, Tobias},
  journal={Health Economics},
  volume={30},
  number={12},
  pages={2956--2973},
  year={2021},
  publisher={Wiley Online Library}
}

@article{bower2005stepped,
  title={Stepped care in psychological therapies: access, effectiveness and efficiency: narrative literature review},
  author={Bower, Peter and Gilbody, Simon},
  journal={The British Journal of Psychiatry},
  volume={186},
  number={1},
  pages={11--17},
  year={2005},
  publisher={Cambridge University Press}
}

@article{nahum2016just,
  title={Just-in-time adaptive interventions (JITAIs) in mobile health: key components and design principles for ongoing health behavior support},
  author={Nahum-Shani, Inbal and Smith, Shawna N and Spring, Bonnie J and Collins, Linda M and Witkiewitz, Katie and Tewari, Ambuj and Murphy, Susan A},
  journal={Annals of behavioral medicine},
  pages={1--17},
  year={2016},
  publisher={Springer}
}

@article{mohr2017personal,
  title={Personal sensing: understanding mental health using ubiquitous sensors and machine learning},
  author={Mohr, David C and Zhang, Mi and Schueller, Stephen M},
  journal={Annual review of clinical psychology},
  volume={13},
  pages={23--47},
  year={2017},
  publisher={Annual Reviews}
}

@article{bauer2020smartphones,
  title={Smartphones in mental health: a critical review of background issues, current status and future concerns},
  author={Bauer, Michael and Glenn, Tasha and Geddes, John and Gitlin, Michael and Grof, Paul and Kessing, Lars V and Monteith, Scott and Faurholt-Jepsen, Maria and Severus, Emanuel and Whybrow, Peter C},
  journal={International journal of bipolar disorders},
  volume={8},
  number={1},
  pages={2},
  year={2020},
  publisher={Springer}
}

@article{rubanovich2017health,
  title={Health app use among individuals with symptoms of depression and anxiety: a survey study with thematic coding},
  author={Rubanovich, Caryn Kseniya and Mohr, David C and Schueller, Stephen M},
  journal={JMIR mental health},
  volume={4},
  number={2},
  pages={e7603},
  year={2017},
  publisher={JMIR Publications Inc., Toronto, Canada}
}

@article{cohen2016integrating,
  title={Integrating patient-generated health data into clinical care settings or clinical decision-making: lessons learned from project healthdesign},
  author={Cohen, Deborah J and Keller, Sara R and Hayes, Gillian R and Dorr, David A and Ash, Joan S and Sittig, Dean F},
  journal={JMIR human factors},
  volume={3},
  number={2},
  pages={e5919},
  year={2016},
  publisher={JMIR Publications Inc., Toronto, Canada}
}

@article{nghiem2023understanding,
  title={Understanding mental health clinicians' perceptions and concerns regarding using passive patient-generated health data for clinical decision-making: qualitative semistructured interview study},
  author={Nghiem, Jodie and Adler, Daniel A and Estrin, Deborah and Livesey, Cecilia and Choudhury, Tanzeem},
  journal={JMIR formative research},
  volume={7},
  number={1},
  pages={e47380},
  year={2023},
  publisher={JMIR Publications Inc., Toronto, Canada}
}

@article{simon2022skating,
  title={Skating the line between general wellness products and regulated devices: strategies and implications},
  author={Simon, David A and Shachar, Carmel and Cohen, I Glenn},
  journal={Journal of Law and the Biosciences},
  volume={9},
  number={2},
  pages={lsac015},
  year={2022},
  publisher={Oxford University Press}
}

@article{lamonica2022informing,
  title={Informing the future of integrated digital and clinical mental health care: synthesis of the outcomes from project synergy},
  author={LaMonica, Haley M and Iorfino, Frank and Lee, Grace Yeeun and Piper, Sarah and Occhipinti, Jo-An and Davenport, Tracey A and Cross, Shane and Milton, Alyssa and Ospina-Pinillos, Laura and Whittle, Lisa and others},
  journal={JMIR mental health},
  volume={9},
  number={3},
  pages={e33060},
  year={2022},
  publisher={JMIR Publications Inc., Toronto, Canada}
}

@article{iorfino2021using,
  title={Using digital technologies to facilitate care coordination between youth mental health services: a guide for implementation},
  author={Iorfino, Frank and Piper, Sarah E and Prodan, Ante and LaMonica, Haley M and Davenport, Tracey A and Lee, Grace Yeeun and Capon, William and Scott, Elizabeth M and Occhipinti, Jo-An and Hickie, Ian B},
  journal={Frontiers in Health Services},
  volume={1},
  pages={745456},
  year={2021},
  publisher={Frontiers Media SA}
}

@article{Chen2024Digital,
  title={Digital Information Ecosystems in Modern Care Coordination and Patient Care Pathways and the Challenges and Opportunities for AI Solutions},
  author={You Chen and Christoph U Lehmann and Bradley A Malin},
  journal={Journal of Medical Internet Research},
  year={2024},
  volume={26},
  doi={10.2196/60258}
}

@article{Backman2024Platform,
  title={Platform-Based Patient-Clinician Digital Health Interventions for Care Transitions: Scoping Review},
  author={Chantal Backman and Rosie Papp and Aurelie Tonjock Kolle and Stephen R. Papp and S. Visintini and Ana Lúcia Schaefer Ferreira de Mello and Gabriela Marcellino de Melo Lanzoni and Anne Harley},
  journal={Journal of Medical Internet Research},
  year={2024},
  volume={26},
  doi={10.2196/55753}
}

@article{murphy2003optimal,
  title={Optimal dynamic treatment regimes},
  author={Murphy, Susan A},
  journal={Journal of the Royal Statistical Society Series B: Statistical Methodology},
  volume={65},
  number={2},
  pages={331--355},
  year={2003},
  publisher={Oxford University Press}
}

@article{aguilera2024effectiveness,
  title={Effectiveness of a digital health intervention leveraging reinforcement learning: results from the Diabetes and Mental Health Adaptive Notification Tracking and Evaluation (DIAMANTE) randomized clinical trial},
  author={Aguilera, Adrian and Ar{\'e}valo Avalos, Marvyn and Xu, Jing and Chakraborty, Bibhas and Figueroa, Caroline and Garcia, Faviola and Rosales, Karina and Hernandez-Ramos, Rosa and Karr, Chris and Williams, Joseph and others},
  journal={Journal of medical Internet research},
  volume={26},
  pages={e60834},
  year={2024},
  publisher={JMIR Publications Toronto, Canada}
}

@article{poolen2025systematic,
  title={Systematic decision-making can help in ending long-term treatments},
  author={Poolen, F and Verhoeven, J and van Schaik, DJF and Reinders, MJ and van der Wart, MT and Vinkers, CH},
  journal={Tijdschrift voor psychiatrie},
  volume={67},
  number={7},
  pages={403--406},
  year={2025}
}

@inproceedings{DanAdlerCHI22_PsychiatricDrugDiscontinuation,
author = {Jo, Eunkyung and Ryu, Myeonghan and Kenderova, Georgia and So, Samuel and Shapiro, Bryan and Papoutsaki, Alexandra and Epstein, Daniel A.},
title = {Designing Flexible Longitudinal Regimens: Supporting Clinician Planning for Discontinuation of Psychiatric Drugs},
year = {2022},
isbn = {9781450391573},
publisher = {Association for Computing Machinery},
address = {New York, NY, USA},
url = {https://doi.org/10.1145/3491102.3502206},
doi = {10.1145/3491102.3502206},
booktitle = {Proceedings of the 2022 CHI Conference on Human Factors in Computing Systems},
articleno = {352},
numpages = {17},
location = {New Orleans, LA, USA},
series = {CHI '22}
}

@book{najavits2002seeking,
  title={Seeking safety: A treatment manual for PTSD and substance abuse},
  author={Najavits, Lisa},
  year={2002},
  publisher={Guilford Publications}
}

@book{beck2024cognitive,
  title={Cognitive therapy of depression},
  author={Beck, Aaron T and Rush, A John and Shaw, Brian F and Emery, Gary and DeRubeis, Robert J and Hollon, Steven D},
  year={2024},
  publisher={Guilford Publications}
}

@article{cuijpers2019role,
  title={The role of common factors in psychotherapy outcomes},
  author={Cuijpers, Pim and Reijnders, Mirjam and Huibers, Marcus JH},
  journal={Annual review of clinical psychology},
  volume={15},
  number={1},
  pages={207--231},
  year={2019},
  publisher={Annual Reviews}
}

@article{kazdin2007mediators,
  title={Mediators and mechanisms of change in psychotherapy research},
  author={Kazdin, Alan E},
  journal={Annu. Rev. Clin. Psychol.},
  volume={3},
  pages={1--27},
  year={2007},
  publisher={Annual Reviews}
}

@String{Computing = "Computing" }

@String{Springer = "Springer-Verlag" }

@article{pennebaker-journalabpsych1986-expressivewritingtrauma,
  title={Confronting a traumatic event: toward an understanding of inhibition and disease.},
  author={Pennebaker, James W and Beall, Sandra K},
  journal={Journal of abnormal psychology},
  volume={95},
  number={3},
  pages={274},
  year={1986},
  publisher={American Psychological Association}
}

@inproceedings{lee2022coauthor,
  title={Coauthor: Designing a human-ai collaborative writing dataset for exploring language model capabilities},
  author={Lee, Mina and Liang, Percy and Yang, Qian},
  booktitle={Proceedings of the 2022 CHI conference on human factors in computing systems},
  pages={1--19},
  year={2022}
}

@article{tewari-mobilehealth2017-adstointerventions,
  title={From ads to interventions: Contextual bandits in mobile health},
  author={Tewari, Ambuj and Murphy, Susan A},
  journal={Mobile health: sensors, analytic methods, and applications},
  pages={495--517},
  year={2017},
  publisher={Springer}
}

@article{king-issuesineducation2006-ejournaling,
  title={E-journaling: A strategy to support student reflection and understanding},
  author={King, Frederick B and LaRocco, Diana},
  journal={Current Issues in Education},
  volume={9},
  year={2006}
}

@article{yu2021reinforcement,
  title={Reinforcement learning in healthcare: A survey},
  author={Yu, Chao and Liu, Jiming and Nemati, Shamim and Yin, Guosheng},
  journal={ACM Computing Surveys (CSUR)},
  volume={55},
  number={1},
  pages={1--36},
  year={2021},
  publisher={ACM New York, NY}
}

@inproceedings{daskalova-chi2021-selfe,
  title={Self-e: Smartphone-supported guidance for customizable self-experimentation},
  author={Daskalova, Nediyana and Kyi, Eindra and Ouyang, Kevin and Borem, Arthur and Chen, Sally and Park, Sung Hyun and Nugent, Nicole and Huang, Jeff},
  booktitle={Proceedings of the 2021 CHI Conference on Human Factors in Computing Systems},
  pages={1--13},
  year={2021}
}

@article{yom2017encouraging,
  title={Encouraging physical activity in patients with diabetes: intervention using a reinforcement learning system},
  author={Yom-Tov, Elad and Feraru, Guy and Kozdoba, Mark and Mannor, Shie and Tennenholtz, Moshe and Hochberg, Irit},
  journal={Journal of medical Internet research},
  volume={19},
  number={10},
  pages={e338},
  year={2017},
  publisher={JMIR Publications Toronto, Canada}
}

@inproceedings{kovacs2021not,
  title={Not now, ask later: users weaken their behavior change regimen over time, but expect to re-strengthen it imminently},
  author={Kovacs, Geza and Wu, Zhengxuan and Bernstein, Michael S},
  booktitle={Proceedings of the 2021 CHI Conference on Human Factors in Computing Systems},
  pages={1--14},
  year={2021}
}

@article{shedler2010efficacy,
  title={The efficacy of psychodynamic psychotherapy.},
  author={Shedler, Jonathan},
  journal={American psychologist},
  volume={65},
  number={2},
  pages={98},
  year={2010},
  publisher={American Psychological Association}
}

@article{fluckiger2017sleeper,
  title={The sleeper effect between psychotherapy orientations: A strategic argument of sustainability of treatment effects at follow-up},
  author={Fl{\"u}ckiger, Christoph and Del Re, AC},
  journal={Epidemiology and psychiatric sciences},
  volume={26},
  number={4},
  pages={442--444},
  year={2017},
  publisher={Cambridge University Press}
}

@article{sbarra2013expressive,
  title={Expressive writing can impede emotional recovery following marital separation},
  author={Sbarra, David A and Boals, Adriel and Mason, Ashley E and Larson, Grace M and Mehl, Matthias R},
  journal={Clinical Psychological Science},
  volume={1},
  number={2},
  pages={120--134},
  year={2013},
  publisher={Sage Publications Sage CA: Los Angeles, CA}
}

@article{hollon2006enduring,
  title={Enduring effects for cognitive behavior therapy in the treatment of depression and anxiety},
  author={Hollon, Steven D and Stewart, Michael O and Strunk, Daniel},
  journal={Annu. Rev. Psychol.},
  volume={57},
  number={1},
  pages={285--315},
  year={2006},
  publisher={Annual Reviews}
}

\end{document}